\documentstyle[11pt,paspconf]{article}

\input epsf 

\def\tcr{t_{\rm cr}}
\def\calR{{\cal R}} 
\def\nb1{{\sf NBODY1} }

\begin{document}

\title{Relaxation of aspherical galaxies}


\author{Christian Boily\altaffilmark{1}}
\affil{Queen Mary \& Westfield College, University of London \\ 
Mile End Road E1 4NS, England.}
\altaffiltext{1}{Now at: Astronomisches Rechen-Institut, 
M\"onchhofstrasse 12-14, Heidelberg D-69120, Germany.}

\begin{abstract} Colour (or, metalicity) gradients are commonly
observed in elliptical galaxies (eg, Danziger et al. 1993). These may
be attributed to stellar evolution processes and gas recycling, or the
conditions at the time the galaxy formed. A tractable approach to
galaxy formation is dissipationless cold collapse. 
This violent but incomplete relaxation 
(eg van Albada 1982)
means that traces of the initial orbit distribution  remain, and so
gradients in galactic properties. Here we seek out 
a possible relation between orbit survival and the on-set of 
radial orbit instabilities (ROIs).  With this in mind, we 
apply Newtonian gravity to collapsing cold spheroids of point masses
as a  model for the Lin-Mestel-Shu (1965) solution for dust.  
   Below  we setup a test problem   which we  
 solve numerically and discuss. 

\end{abstract}



\section{N-body integration} Our toy model consists of a uniform sphere 
containing two mass  components, one inside, the other outside,  
$r = \pi$; the bounding radius $ = 5$. Each mass bin is sampled with 5,000 bodies. 
Making $m_1/m_2 = 1/3$, the mean density is everywhere the same and
hence the 
radial free-fall 
time is unique. With the more massive stars confined to 
  $\pi < r < 5$, the intention is to favour
mass ejection from, or otherwise disturb, the lighter central distribution. 
 This is a robust test of  orbit survival in energy space under violent relaxation. \newline

The 10,000-particle simulation was performed with the code 
\nb1  on HARP computer. The length  $l$ of the smoothed potential $ = 3\times 10^{-4}$ (in system units), well below close-encounter  impact-parameter 
values. The 
simulation ran for a total of $6$  initial crossing time $(\equiv \tcr)$. A small 
velocity was given to each star such that the virial ratio $Q \equiv -T/W = 1/200 $ 
initially. 
Two more simulations were performed with the same random seed distribution, but now flattening the z-axis by  factors of  2 (hamburger) and 10 (pancake): this covers an 
appropriate range of spheroid aspect ratios. 

The two-body relaxation timescale for this system is $\simeq 60\, \tcr$
(Chernoff \& Weinberg 1990; Spitzer 1987) so evolution over $6\, \tcr$
is largely collisionless. 

\section{Discussion} 
We measured orbit-mixing by monitoring in time 
the ranking $\calR_{t,i}$ of each $(i^{th})$
 star relative to one another, introducing 

\[ \Delta \equiv \left( \frac{1}{2N} \sum_{S}\sum_{i_{{\rm bin}}} \left[ \calR_{t,i} - \calR_{o,i} \right]^{2}\right)^{1/2}. \label{c2-eq:Delta} \] 
Completely mixed orbits give  
$ \Delta \simeq 57.7$ when  $S$ the number of rank bins $ = 200$. 
Both  spherical and `pancake' collapses 
erase memory of the initial vertical $(z)$  structure rather well,
however this is not so 
 for the `hamburger' simulation: we find post-collapse 
 $\Delta_z \approx 42$, well below the expected value for fully-mixed
systems. 
Inspection of the principal components of the inertia tensor
showed that both mass components relaxed to a triaxial figure, 
with axis ratios $1.00:0.87:0.80$, in rough
 agreement with  values derived from a scaling argument
applied to LMS-type collapses   
(see Boily et al. 1998). The heavy masses
filled a volume twice as large (larger $I_{xx}$ per unit mass) so the
lighter population orbited in the background potential of the heavies.

For the case of the spherical collapse, we found at the bounce and
thereafter that the lighter masses formed a spheroidal mass profile,
while the heavier particles remained spherically distributed.  
Furthermore, they were also (mildly) more 
centrally concentrated. As we let  
this system evolve  up to $50 \, \tcr$, we found that orbit-orbit 
interactions with the heavy population reestablished sphericity 
(see also Theis \& Spurzem 1998): in the process heavy and light stars
moved out- and in-wards respectively, opposite the trend expected 
of two-body relaxation.  Near the end of the run,
 both populations filled the same spherical  volume. 

 These results suggest further
investigations into processes driving ROIs (eg, Aguilar \& Merritt
1990) to establish long-term effects of a  bias in the initial mass 
distribution. 

\section{Acknowledgements}  Warm thanks to  the organisers of the conference,
and Monica Valluri in particular, for agreeing 
to this contribution {\it in absentia}.

\end{document}


\title{Application of a parallel hybrid $N$--body SCF code}
\author{Marc Hemsendorf}
\affil{Astronomisches Rechen-Institut,
M\"onchhofstr.~12--14, 69120 Heidelberg}

\section{Introduction}
The stellar dynamics of galactic nuclei are in the focus of
astronomical interest. Observations show that many if not all
galactic centres harbor supermassive black holes. Their mass
determination, however, relies on stationary modelling,
which does not provide a self-consistent picture of the evolution
of galactic nuclei.

Another point of interest is the behavior of binary black holes
within a dense central stellar cluster. To what extend does star gas or
dynamical friction influence the results? These applications
need a high precision collisional code, which can handle a great
number of particles (Quinlan \& Hernquist 1997).

The $N^2$ dependency of computational time restricts high precision
direct $N$--body calculations to a comparatively small number of
particles.  In contrast, codes which calculate the mean potential of
particle distribution do not have this problem but are not sufficient
to treat the dynamics of central stellar clusters. Since many stellar
systems show a clear core halo structure, one can profit from the
advantages of both approaches in a hybrid code.  The concepts of such
a hybrid scheme and its results are described here.

\section{Merging of \emph{NBODY6++} with \emph{SCF}}
The primary $N$--body integrator for collisional systems like globular
clusters is an Aarseth \emph{NBODYx} code (Aarseth 1993). The most
complex one, \emph{NBODY6++}, is highly efficient on serial
computers. This code has been ported to parallel computers (Spurzem
\& Baumgart 1997) where it can show its advantages compared to $N^2$
codes running 
on special purpose computers. Nearly all parts of the code, like the
ones for binary or chain treatment, can run in parallel as long as the
problem size is large enough.

To embed \emph{NBODY6++} in a living, nearly spherical halo, we choose
the \emph{SCF} method by Clutton--Brock (1973) and Hernquist \&
Ostriker (1992). The parametrised $\alpha$--law basis functions
published by Zhao (1996), which are used in this code instead, provide
a much higher flexibility. The Hermite integration scheme in
\emph{NBODY6++} needs the input of the forces $\mathbf{a}_i$ and force
derivatives $\mathbf{\dot{a}}_i$. Unlike in other mean field methods,
these can be added to SCF analytically.

The angular momentum of the particles is the criterion to distribute
the work over both codes. Particles with a low angular momentum are
more reactive to their neighbors and therefore given to
\emph{NBODY6++}. In a Plummer potential, this criterion assigns all
spatially central particles to direct integrator.

The \emph{SCF} method relies on an analytically given distribution
function for density and for potential. Thus, a cut off potential is
very hard to handle.  This problem can be solved by calculating the
the forces of the whole particle set using \emph{SCF}. \emph{NBODY6++}
and \emph{SCF} re--calculate the forces in the subset below the
critical angular momentum. In this process the direct code takes the
difference force of the two field method runs as an external
potential. The time steps of the field method are usually longer than
the ones of \emph{NBODY6++}.

\section{Testing}
One important test of the code is the benchmarking. The table below
shows the computation time in seconds for 40 SCF time steps for various
numbers of processing elements. About half the number of particles
were below the critical angular momentum.

\begin{center}
\begin{tabular}{c|ccccc}
$N_{\rm tot}$ & 16 & 32 & 64 & 128 & 256 \\ \hline
16384 & 22.3 & 12.4 & 8.3 & 7.8 & 10.2 \\
32768 & 71.8 & 38.5 & 24.9 & 19.6 & 34.1 \\
65536 & 266.4 & 147.8 & 81.2 & 64.5 & 59.7
\end{tabular}
\end{center}

One realization of a Plummer model was evolved by the hybrid code and a
HARP version of \emph{NBODY6++} for several crossing
times in order to check the difference in relaxation effects.
In the field method part of the hybrid code one observes less
relaxation than what the direct method finds. Introducing Spitzer
relaxation to \emph{SCF}, however, might be a way to tackle this problem.

\acknowledgments
I like to thank my co--workers R.~Spurzem and S.~Sigurdsson who started
this project, S.~Aarseth, D.~Heggie, W.~Sweatman and G.~Hensler for
fruitful help and discussion. This project is funded by \emph{Deutsche
Forschungsgemeinschaft} (DFG) project Sp 345/9-1. Technical help and
computer resources are provided by \emph{HLRS} in Stuttgart,
\emph{HLRZ} in J\"ulich, \emph{EPCC} in Edinburgh, \emph{ZIB} in Berlin,
University of Heidelberg and University of Kiel.


\title{Simulations of star clusters and Galactic Nuclei: rotation and
other challenges}
\author{R. Spurzem and C. Einsel}
\affil{Astronomisches Rechen-Institut, M\"onchhofstra\ss e 12-14,
D-69120 Heidelberg, Germany}

\section{Introduction}

With strong efforts -- both observationally and theoretically -- it has now
become possible not only to find evidence for the existence of black holes in
the centres of galaxies, but also to set up an expression for the
mass distribution of black holes with respect to the mass of the parent
galaxy - or its spheroid component (Kormendy \& Richstone 1995). 
In deriving
quantities of central objects from the orbit structure of the surrounding
star clusters it is crucial whether 2- or 3-integral models are made and
which integrals are chosen. Using $E$ and $J_{z}$ van der Marel {\em et al.}
(1998) analyze the central cluster of M32 and find significant amount
of rotation around a black hole mass of roughly
$M_{bh}=3.4\times 10^{6}M_{\odot}$. The relaxation time at least in this system
is several times less than its age, so that corresponding processes may play
an important role in the evolution and build-up of central dark objects and
their surrounding clusters.

Only recently first efforts are being started to
model rotating galactic nuclei
including the effects of stellar two-body relaxation, star accretion
onto the black hole, movement and growth of one or two central black holes.
There is an $N$-body approach by Arabadjis \& Richstone (1998ab) using
a TREE-code.
We have undertaken an effort to use
a simplified (2 integrals of motion, equal masses)
2D-Fokker-Planck model to follow the evolution of self-gravitating,
rotating star clusters into core collapse without central mass
(Einsel \& Spurzem 1998). The gravo-gyro catastrophe, earlier postulated
by Hachisu (1979) by using a linear stability analysis, has been
rediscovered, followed by a self-similar contraction phase.

Now, as a second step, we study multi-mass star clusters. The well-known
mass segregation instability is now accompanied by the gravo-gyro and
subsequent effects governed by the interplay between rotation and
two--body relaxation. 

In the Figure the evolution of the central angular velocity of the
different stellar masses (indicated at the lines) is plotted as a
function of total central density (which can be considered as a
scaled time). The initial model was a Michie-King like model with $W_0=3$,
dimensionless rotational parameter $\omega_0=0.30$ and a Salpeter
mass spectrum. We observe that a fast rotating core of the heavy masses
is built up, and conclude that the hitherto not carefully analysed
effect of rotation in relaxing star clusters introduces unexpected and
very interesting kinematic features. 

\begin{center}
\begin{figure}[tb]
\psfig{file=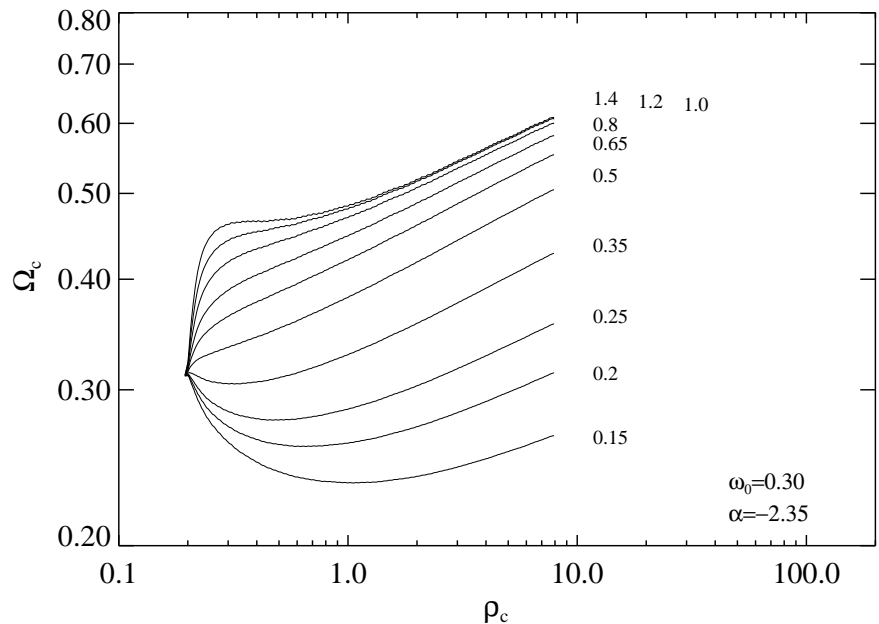,width=120mm,height=85mm,angle=90}


\caption{ Evolution of central angular velocities versus centraldensity of all 10 mass bins
used in the calculation for an initial model with $W_{0}=3$ and
$\omega_0=0.90$. The highest mass bin with $m=1.5~M_{\odot}$ is the
uppermost curve, while the lowest mass bin with $m=0.15~M_{\odot}$ can
be found as the lowest curve.
}

\end{figure}
\end{center}

This models will be published in more details elsewhere (Einsel \& Spurzem
1999), and the inclusion of a central massive star-accreting black
hole is in planned. 

\acknowledgments

Financial support by German Science Foundation (DFG, grant no. Sp 345/3-3)
is gratefully acknowledged, as well as computing time provided by
H\"ochstleistungsrechenzentrum (HLRZ) J\"ulich on a CRAY T90 vector computer.